\documentclass{bmcart}

\usepackage{amsthm,amsmath}
\usepackage{bm}
\usepackage{url}
\usepackage{dsfont}
\usepackage{graphicx}
\usepackage[utf8]{inputenc} 

\startlocaldefs
\endlocaldefs

\DeclareMathOperator{\logit}{logit}
\DeclareMathOperator{\expit}{expit}

\begin{document}

\begin{frontmatter}

\begin{fmbox}
\dochead{Research}


\title{Two-stage matching-adjusted indirect comparison}


\author[
   addressref={aff1,aff2},                   
   corref={aff1,aff2},                       
   email={antonio.remiro-azocar@bayer.com}   
]{\inits{ARA}\fnm{Antonio} \snm{Remiro-Az\'ocar}}


\address[id=aff1]{
  \orgname{Medical Affairs Statistics, Bayer plc}, 
  \street{400 South Oak Way},                     %
  \city{Reading},                              
  \cny{UK}                                    
}
\address[id=aff2]{%
  \orgname{Department of Statistical Science, University College London},
  \street{1-19 Torrington Place},
  \city{London},
  \cny{UK}
}



\end{fmbox}


\begin{abstractbox}

\begin{abstract} 
\textbf{Background:} Anchored covariate-adjusted indirect comparisons inform reimbursement decisions where there are no head-to-head trials between the treatments of interest, there is a common comparator arm shared by the studies, and there are patient-level data limitations. Matching-adjusted indirect comparison (MAIC), based on propensity score weighting, is the most widely used covariate-adjusted indirect comparison method in health technology assessment. MAIC has poor precision and is inefficient when the effective sample size after weighting is small.

\noindent \textbf{Methods:} A modular extension to MAIC, termed two-stage matching-adjusted indirect comparison (2SMAIC), is proposed. This uses two parametric models. One estimates the treatment assignment mechanism in the study with individual patient data (IPD), the other estimates the trial assignment mechanism. The first model produces inverse probability weights that are combined with the odds weights produced by the second model. The resulting weights seek to balance covariates between treatment arms and across studies. A simulation study provides proof-of-principle in an indirect comparison performed across two randomized trials. Nevertheless, 2SMAIC can be applied in situations where the IPD trial is observational, by including potential confounders in the treatment assignment model. The simulation study also explores the use of weight truncation in combination with MAIC for the first time.  

\noindent \textbf{Results:} Despite enforcing randomization and knowing the true treatment assignment mechanism in the IPD trial, 2SMAIC yields improved precision and efficiency with respect to MAIC in all scenarios, while maintaining similarly low levels of bias. The two-stage approach is effective when sample sizes in the IPD trial are low, as it controls for chance imbalances in prognostic baseline covariates between study arms. It is not as effective when overlap between the trials' target populations is poor and the extremity of the weights is high. In these scenarios, truncation leads to substantial precision and efficiency gains but induces considerable bias. The combination of a two-stage approach with truncation produces the highest precision and efficiency improvements. 

\noindent \textbf{Conclusions:} Two-stage approaches to MAIC can increase precision and efficiency with respect to the standard approach by adjusting for empirical imbalances in prognostic covariates in the IPD trial. Further modules could be incorporated for additional variance reduction or to account for missingness and non-compliance in the IPD trial. 
\end{abstract}


\begin{keyword}
\kwd{health technology assessment}
\kwd{indirect treatment comparison}
\kwd{matching-adjusted indirect comparison}
\kwd{covariate adjustment}
\kwd{covariate balance}
\kwd{inverse probability of treatment weighting}
\kwd{evidence synthesis}
\end{keyword}


\end{abstractbox}
%

\end{frontmatter}



\section*{Background}\label{sec1}

In many countries, health technology assessment (HTA) addresses whether new treatments should be reimbursed by public health care systems \cite{vreman2020decision}. This often requires estimating relative effects for interventions that have not been directly compared in a head-to-head trial \cite{sutton2008use}. Consider that there are two active treatments of interest, say $A$ and $B$, that have not been evaluated in the same study, but have been contrasted against a comparator $C$ in different studies. In this situation, an indirect comparison of relative treatment effect estimates is required. The analysis is said to be \textit{anchored} by the common comparator $C$.

A typical situation in HTA is that where a pharmaceutical company has individual patient data (IPD) from its own study comparing $A$ versus $C$, which we shall denote the \textit{index} trial, but only published aggregate-level data (ALD) from another study comparing $B$ versus $C$, which we call the \textit{competitor} trial. In this two-study scenario, cross-trial imbalances in effect measure modifiers, effect modifiers for short, make the standard indirect treatment comparisons \cite{bucher1997results} vulnerable to bias \cite{dias2013evidence}. Novel covariate-adjusted indirect comparison methods have been introduced to account for these imbalances and provide equipoise to the comparison \cite{phillippo2016nice, phillippo2018methods, remiro2021methods, remiro2021conflating, remiro2021effect}. 

The most popular methodology \cite{phillippo2019population} in peer-reviewed publications and submissions for reimbursement is matching-adjusted indirect comparison (MAIC) \cite{signorovitch2010comparative, signorovitch2012comparative, signorovitch2012matching}. MAIC weights the subjects in the index trial to create a ``pseudo-sample'' with balanced moments with respect to the competitor trial. The standard formulation of MAIC proposed by Signorovitch et al. \cite{signorovitch2010comparative} uses a method of moments to estimate a logistic regression, which models the trial assignment mechanism. The weights are derived from the fitted model and represent the odds of assignment to the competitor trial for the subjects in the IPD, conditional on selected baseline covariates.

Under no failures of assumptions, MAIC has produced unbiased treatment effect estimation in simulation studies \cite{remiro2021methods, hatswell2020effects, cheng2020statistical, wang2021matching, petto2019alternative, kuhnast2017evaluation, weber2020comparison, jiang2020performance}. Nevertheless, there are some concerns about its inefficiency and instability, particularly where covariate overlap is poor and effective sample sizes (ESSs) after weighting are small \cite{phillippo2020assessing}. These scenarios are pervasive in health technology appraisals \cite{phillippo2019population}. In these cases, weighting methods are sensitive to inordinate influence by a few subjects with extreme weights and are vulnerable to poor precision. A related concern is that feasible numerical solutions may not exist where there is no common covariate support \cite{phillippo2020assessing,jackson2020alternative}. Where overlap is weak, methods based on modeling the outcome expectation exhibit greater precision and efficiency than MAIC \cite{phillippo2020assessing, remiro2021parametric,remiro2020marginalization, phillippo2021target} but are prone to extrapolation, which may lead to severe bias under model misspecification \cite{ho2007matching, rubin1997estimating}. 

Consequently, modifications of MAIC that seek to maximize precision have been presented. An alternative implementation estimates the weights using entropy balancing \cite{petto2019alternative,belger2015inclusion}. The proposal is similar to the standard method of moments, with the additional constraint that the weights are as close as possible to unit weights, potentially penalizing extreme weighting schemes. While the approach has appealing computational properties, Phillippo et al.~have proved that it is mathematically equivalent to the standard method of moments \cite{phillippo2020equivalence}. 

More recently, Jackson et al.~have developed a distinct weight estimation procedure that satisfies the conventional method of moments while explicitly maximizing the ESS \cite{jackson2020alternative}. This translates into minimizing the dispersion of the weights, with more stable weights improving precision at the expense of inducing bias. 

Other approaches to limit the undue impact of extreme weights involve truncating or capping the weights. These are common in survey sampling \cite{elliott2000model} and in many propensity score settings \cite{lee2011weight,cole2008constructing} but are yet to be investigated specifically alongside MAIC. Again, a clear trade-off is involved from a bias-variance standpoint. Lower variance comes at the cost of sacrificing balance and accepting bias \cite{moore2012causal, leger2022causal}. Limitations of weight truncation are that it shifts the target population or estimand definition, and that it requires arbitrary ad hoc decisions on cutoff thresholds. 

In order to gain efficiency, I propose a modular extension to MAIC which uses two parametric models. One estimates the treatment assignment mechanism in the index study, the other estimates the trial assignment mechanism. The first model produces inverse probability of treatment weights that are combined with the weights produced by the second model. I term this approach \textit{two-stage matching-adjusted indirect comparison} (2SMAIC). 

In the anchored scenario, the conventional version of MAIC relies on randomization in the index trial. In this setting, the treatment assignment mechanism (the true conditional probability of treatment among the subjects) is typically known. In addition, randomization ensures that there is no confounding on expectation. Therefore, it may seem counter-intuitive to model the treatment assignment mechanism in this study. Nevertheless, this additional step is beneficial to control for finite-sample imbalances in prognostic baseline covariates. These imbalances often arise due to chance and correcting for them leads to efficiency gains. 

An advantage of 2SMAIC is that, due to incorporating a treatment assignment model, it is also applicable where the index study is observational. In this case, within-study randomization is not leveraged and concerns about internal validity must be addressed by including potential confounders of the treatment-outcome association in the treatment assignment model. The estimation procedure for the trial assignment weights does not necessarily need to be that of Signorovitch et al. \cite{signorovitch2010comparative} and alternative methods could be used \cite{wang2021matching, jackson2020alternative}. Further modules could be incorporated to account for missingness \cite{seaman2013review} and non-compliance \cite{cain2009inverse}, e.g.~dropout or treatment switching, in the index trial. 

I conduct a proof-of-concept simulation study to examine the finite-sample performance of 2SMAIC with respect to the standard MAIC when the index study is an RCT. The two-stage approach improves the precision and efficiency of MAIC without introducing bias. The results are consistent with previous research on the efficiency of propensity score estimators \cite{lunceford2004stratification, hahn1998role}. Finally, the use of weight truncation in combination with MAIC is explored for the first time. Example code to implement the methodologies in \texttt{R} is provided in Additional file 1. 

\section*{Methods}\label{sec2}

\subsection*{Context and data structure}\label{subsec21}

We focus on the following setting, which is common in submissions to HTA agencies. Let $S$ and $T$ denote indicators for the assigned study and the assigned treatment, respectively. There are two separate studies that enrolled distinct sets of participants and have now been completed. The index study ($S=1$) compares active treatment $A$ ($T=1$) versus $C$ ($T=0$), e.g.~standard of care or placebo. The competitor study ($S=2$) evaluates active treatment $B$ ($T=2$) versus $C$ ($T=0$). Covariate-adjusted indirect comparisons such as MAIC perform a treatment comparison in the $S=2$ sample, implicitly assumed to be of policy interest. We ask ourselves the question: what would be the marginal treatment effect for $A$ versus $B$ had these treatments been compared in an RCT conducted in $S=2$? 

The marginal treatment effect for $A$ vs.~$B$ is estimated on the linear predictor (e.g.~mean difference, log-odds ratio or log hazard ratio) scale as:

\begin{equation}
\hat{\Delta}_{12}^{(2)} = \hat{\Delta}_{10}^{(2)} - \hat{\Delta}_{20}^{(2)},
\label{equation1}
\end{equation}
where $\hat{\Delta}_{10}^{(2)}$ is an estimate of the hypothetical marginal treatment effect for $A$ vs.~$C$ in the competitor study sample, and $\hat{\Delta}_{20}^{(2)}$ is an estimate of the marginal treatment effect of $B$ vs.~$C$ in the competitor study sample. MAIC uses weighting to transport inferences for the marginal $A$ vs.~$C$ treatment effect from $S=1$ to $S=2$. The estimate $\hat{\Delta}_{10}^{(2)}$ is produced, which is then input into Equation \ref{equation1}. Because the within-trial relative effect estimates are assumed statistically independent, their variances are summed to estimate the variance of the marginal treatment effect for $A$ vs.~$B$.  

The manufacturer submitting evidence for reimbursement has access to individual-level data $\mathcal{D}_{AC}=({\bm{x},\bm{t},\bm{y}})$ on covariates, treatment and outcomes for the participants in its trial. Here, $\bm{x}$ is a matrix of pre-treatment baseline covariates (e.g.~comorbidities, age, gender), of size $n \times k$, where $n$ is the total number of subjects in the study sample and $k$ is the number of covariates. A row vector $\bm{x}_i=  ( x_{i,1}, x_{i,2}, \dots, x_{1,k} )$ of $k$ covariates is recorded for each participant $i=1,\dots n$. We let $\bm{y}=(y_1, y_2, \dots, y_n)$ denote a vector of the clinical outcome of interest and $\bm{t}=(t_1, t_2, \dots, t_n)$ denote a binary treatment indicator vector. We shall assume that there is no loss to follow-up or missing data on covariates, treatment and outcome in $\mathcal{D}_{AC}$.

We consider all baseline covariates to be prognostic of the clinical outcome and select a subset of these, $\bm{z} \subseteq \bm{x}$, as marginal effect modifiers for $A$ with respect to $C$ on the linear predictor scale, with a row vector $\bm{z}_i$ recorded for each patient $i$. In the absence of randomization, the variables in $\bm{x}$ would induce confounding between the treatment arms in the index study (internal validity bias). On the other hand, cross-trial imbalances in the variables in $\bm{z}$ induce external validity bias with respect to the competitor study sample. 

Neither the manufacturer submitting the evidence nor the HTA agency evaluating it have access to IPD for the competitor trial. We let $\mathcal{D}_{BC}=[\bm{\theta_x}, \hat{\Delta}_{20}^{(2)}, \hat{V}(\hat{\Delta}_{20}^{(2)})]$ represent the published ALD that is available for this study. No patient-level covariates, treatment or outcomes are available. Here, $\bm{\theta_x}$ denotes a vector of means or proportions for the covariates; although higher-order moments such as variances may also be available. An assumption is that a sufficiently rich set of baseline covariates has been measured for the competitor study. Namely, that summaries for the subset $\bm{\theta_z} \subseteq \bm{\theta_x}$ of covariates that are marginal effect modifiers are described in the table of baseline characteristics in the study publication.

Also available is an internally valid estimate $\hat{\Delta}_{20}^{(2)}$ of the marginal treatment effect for $B$ vs.~$C$ in the competitor study sample, and an estimate $\hat{V}(\hat{\Delta}_{20}^{(2)})$ of its variance. These are either directly reported in the publication or, assuming that the competitor study is a well-conducted RCT, derived from crude aggregate outcomes in the literature. 

\subsection*{Matching-adjusted indirect comparison}\label{C2S2}

In MAIC, IPD from the index study are weighted so that the moments of selected covariates are balanced with respect to the published moments of the competitor study. The weight $w_i$ for each participant $i$ in the index trial is estimated using a logistic regression:
\begin{equation}
\ln(w_i) = 
\ln[w(\bm{z}_i)]
= \ln  \Bigg [ \frac{Pr(S=2 \mid \bm{z}_i)}{1 - Pr(S=2 \mid \bm{z}_i)} \Bigg ]
=
\alpha_0 + \bm{z}_i\bm{\alpha_1}, 
\label{equation2}
\end{equation}
where $\alpha_0$ is the model intercept and $\bm{\alpha_1}$ is a vector of model coefficients. While most applications of weighting, e.g.~to control for confounding in observational studies, construct ``inverse probability'' weights for \textit{treatment} assignment, MAIC uses ``odds weighting'' \cite{westreich2017transportability, dahabreh2020extending} to model \textit{trial} assignment. The weight $w_i$ represents the conditional odds that an individual $i$ with covariates $\bm{z}_i$, selected as marginal effect modifiers, is enrolled in the competitor study. Alternatively, the weight represents the inverse conditional odds that the individual is enrolled in the index study.

The logistic regression parameters in Equation \ref{equation2} cannot be derived using conventional methods such as maximum-likelihood estimation, due to unavailable IPD for the competitor trial. Signorovitch et al.~propose using a method of moments instead to enforce covariate balance across studies \cite{signorovitch2010comparative}. Prior to balancing, the IPD covariates are centered on the means or proportions published for the competitor trial. The centered covariates for subject $i$ in the IPD are defined as $\bm{z^*}_i = \bm{z}_i - \bm{\theta_z}$. 

Weight estimation involves minimizing the objective function:
\begin{equation}
Q(\bm{\alpha_1}) = \sum_{i=1}^{n} \exp \left ( \boldsymbol{z^*}_i \boldsymbol{\alpha_1}\right ).
\label{equation3}
\end{equation}
The function $Q(\bm{\alpha_1})$ is convex \cite{signorovitch2010comparative} and can be minimized using standard convex optimization algorithms \cite{nocedal2006numerical}. Provided that there is adequate overlap, minimization yields the unique finite solution: $\hat{\bm{\alpha}}_{\bm{1}}=\textnormal{argmin}[Q(\bm{\alpha_1})]$. Feasible solutions do not exist if all the values observed for a covariate in $\bm{z}$ are greater or lesser than its corresponding element in $\bm{\theta_z}$ \cite{jackson2020alternative}.

After minimizing the objective function in Equation \ref{equation3}, the weight estimated for the $i$-th participant in the IPD is:
\begin{equation}
\hat{w}_i = \exp(\boldsymbol{z^*}_i\boldsymbol{\hat{\alpha}}_{\boldsymbol{1}}).
\label{equation4}
\end{equation}
The estimated weights are relative, in the sense that any weights that are proportional are equally valid \cite{jackson2020alternative}. Weighting reduces the ESS of the index trial. The approximate ESS after weighting is typically estimated as $\left(\sum_i^n\hat{w}_i\right)^2/\sum_i^n\hat{w}_i^2$ \cite{phillippo2016nice, kish1965survey}. Low values of the ESS suggest that a few influential participants with disproportionate weights dominate the reweighted sample. 

Consequently, marginal mean outcomes for treatments $A$ and $C$ in the competitor study sample ($S=2$) are estimated as the weighted average:
\begin{equation}
\hat{\mu}^{(2)}_t = \frac{\sum_{i=1}^{n_t} y_{i,t} \hat{w}_{i,t}}{\sum_{i=1}^{n_t} \hat{w}_{i,t}},
\label{equation5}
\end{equation}
where $n_t$ denotes the number of participants assigned to treatment $t \in \{0, 1\}$ of the index trial, $y_{i,t}$ represents the observed clinical outcome for subject $i$ in arm $t$, and $\hat{w}_{i,t}$ is the weight assigned to patient $i$ under treatment $t$. For binary outcomes, $\hat{\mu}_t$ would estimate the expected marginal outcome probability under treatment $t$. Absolute outcome estimates may be desirable as inputs to health economic models \cite{phillippo2021target} or in unanchored comparisons made in the absence of a common control group. 

In anchored comparisons, the objective is to estimate a relative effect for $A$ vs.~$C$, as opposed to absolute outcomes. Indirect treatment comparisons are typically conducted on the linear predictor scale \cite{bucher1997results, dias2013evidence, phillippo2018methods}. Consequently, this scale is also used to define effect modification, which is scale specific \cite{phillippo2016nice}. 

One can convert the mean absolute outcome predictions produced by Equation \ref{equation5} from the natural scale to the linear predictor scale, and compute the marginal treatment effect for $A$ vs.~$C$ in $S=2$ as the difference between the average linear predictions:
\begin{equation}
\hat{\Delta}_{10}^{(2)} =  g \big( \hat{\mu}_1^{(2)} \big ) - g \big ( \hat{\mu}_0^{(2)} \big ).
\label{equation6}
\end{equation}
Here, $g(\cdot)$ is an appropriate link function, e.g.~the identity link produces a mean difference for continuous-valued outcomes, and the $\logit \big (\hat{\mu}^{(2)}_t \big) = \ln \big [\hat{\mu}^{(2)}_t/\big(1-\hat{\mu}^{(2)}_t \big)\big]$ generates a log-odds ratio for binary outcomes. Different, potentially more interpretable, choices such as relative risks and risk differences are possible for the marginal contrast. One can map to these scales by manipulating $\hat{\mu}_1^{(2)}$ and $\hat{\mu}_0^{(2)}$ differently. 

Alternatively, the weights generated by Equation \ref{equation4} can be used to fit a simple regression of outcome on treatment to the IPD \cite{schafer2008average}. The model can be fitted using maximum-likelihood estimation, weighting the contribution of each individual $i$ to the likelihood by $\hat{w}_i$. In this approach, the treatment coefficient of the fitted weighted model is the estimated marginal treatment effect $\hat{\Delta}_{10}^{(2)}$ for $A$ vs.~$C$ in $S=2$. 

The original approach to MAIC uses a robust sandwich-type variance estimator \cite{robins2000marginal} to compute the standard error of $\hat{\Delta}_{10}^{(2)}$. This relies on large-sample properties and has understated variability with small ESSs in a previous simulation study investigating MAIC \cite{remiro2021methods} and in other settings \cite{fay2001small, chen2017variance, tipton2017implications, raad2020evaluation}. In addition, most implementations of the sandwich estimator, e.g.~when fitting the weighted regression \cite{zeileis2006object}, ignore the estimation of the trial assignment model, assuming the weights to be fixed quantities. While analytic expressions that incorporate the estimation of the weights could be derived, a practical alternative is to resample via the ordinary non-parametric bootstrap \cite{remiro2021parametric, efron1994introduction, sikirica2013comparative}, re-estimating the weights and the marginal treatment effect for $A$ vs.~$C$ in each bootstrap iteration. Point estimates, standard errors and interval estimates can be directly calculated from the bootstrap replicates. 

We briefly describe the assumptions required by MAIC and their implications: 
\begin{enumerate}
    \item \textbf{Internal validity} of the effect estimates derived from the index and competitor studies. This is certainly feasible where the studies are RCTs because randomization ensures exchangeability over treatment assignment on expectation. While internal validity may hold in RCTs, it is a more stringent condition for observational studies. The absence of informative measurement error, missing data, non-adherence, etc.~is assumed. 
    \item \textbf{Consistency under parallel studies} \cite{hartman2015sample}. There is only one well-defined version of each treatment \cite{rubin1980randomization} or any variations in the versions of treatment are irrelevant \cite{vanderweele2013causal, vanderweele2009concerning}. This applies to the common comparator $C$ in particular. 
    \item \textbf{Conditional transportability} (exchangeability) of the marginal treatment effect for $A$ vs.~$C$ from the index to the competitor study \cite{westreich2017transportability}. Namely, trial assignment does not affect this measure, conditional on $\bm{z}$. Prior research has referred to this assumption as the \textit{conditional constancy of relative effects} \cite{phillippo2016nice, phillippo2018methods, remiro2021effect}. It is plausible if $\bm{z}$ comprises all of the covariates that are considered to modify the marginal treatment effect for $A$ vs.~$C$ (i.e., there are no unmeasured effect modifiers) \cite{hernan2011compound, o2014generalizing, zhang2016new}.\footnote{This assumption is strong and untestable. Nevertheless, it is weaker than that required by unanchored comparisons. Unanchored comparisons compare absolute outcome means as opposed to relative effect estimates. Therefore, these rely on the conditional exchangeability of the absolute outcome mean under active treatment (conditional constancy of absolute effects) \cite{phillippo2016nice, phillippo2018methods, dahabreh2020extending, rudolph2017robust}. This requires capturing all factors that are prognostic of outcome given active treatment.} 
    \item \textbf{Sufficient overlap}. The ranges of the selected covariates in $S=1$ should cover their respective moments in $S=2$. Overlap violations can be deterministic or random. The former arise structurally, due to non-overlapping trial target populations (eligibility criteria). The latter arise empirically due to chance, particularly where sample sizes are small \cite{westreich2010invited}. Therefore, overlap can be assessed based on absolute sample sizes. The ESS is a convenient one-number diagnostic.
    \item \textbf{Correct specification of the $S=2$ covariate distribution}. Analysts can only approximate the joint distribution because IPD are unavailable for the competitor study. Covariate correlations are rarely published for $S=2$ and therefore cannot be balanced by MAIC. In that case, they are assumed equal to those in the pseudo-sample formed by weighting the IPD \cite{phillippo2016nice}.
\end{enumerate}
I make a brief remark on the specification of the parametric trial assignment model in Equation \ref{equation2}. This does not necessarily need to be correct as long as it balances all the covariates, and potential transformations of these covariates, e.g.~polynomial transformations and product terms, that modify the marginal treatment effect for $A$ vs.~$C$ \cite{remiro2021effect, remiro2021parametric}. Squared terms are often included to balance variances for continuous covariates \cite{signorovitch2010comparative} but initial simulation studies do not report performance benefits \cite{hatswell2020effects, petto2019alternative}. This is probably due to greater reductions in ESS and precision \cite{phillippo2021target}.

The identification of effect modifiers will likely require prior background knowledge and substantive domain expertise. Bias-variance trade-offs are also important. Failing to include an influential effect modifier in $\bm{z}$, whether in imbalance or not, leads to bias in $S=2$ \cite{phillippo2016nice, dahabreh2020extending, stuart2010matching}. On the other hand, the inclusion of covariates that are not effect modifiers reduces overlap, thereby increasing the chance of extreme weights. This decreases precision without improving the potential for bias reduction \cite{phillippo2018methods, nie2013likelihood}, even if the covariates are strongly imbalanced across studies. That is, even if they predict or are associated to trial assignment. 

Put simply, as is the case for other weighting-based methods \cite{brookhart2006variable, shortreed2017outcome}, MAIC is potentially unbiased if either the trial assignment mechanism or the outcome-generating mechanism is known, with the latter leading to better performance due to reduced variance and increased efficiency. 

\subsection*{Two-stage matching-adjusted indirect comparison}\label{C2S3}

While the standard MAIC models the trial assignment mechanism, two-stage MAIC (2SMAIC) additionally models the treatment assignment mechanism in the index trial. The treatment assignment model is estimated to produce inverse probability of treatment weights. Then, these are combined with the odds weights generated by the standard MAIC. The resulting weights seek to balance covariate moments between the studies and the treatment arms of the index trial.  

For the treatment assignment mechanism, a propensity score logistic regression of treatment on the covariates is fitted to the IPD:
\begin{equation}
\logit[e_i] = \logit[e(\bm{x}_i)] =
\logit[Pr(T=1\mid \bm{x}_i)] = \beta_0 + \bm{x}_i \bm{\beta_1},
\label{equation7}
\end{equation}
where $\beta_0$ and $\bm{\beta_1}$ parametrize the logistic regression. The propensity score $e_i$ is defined as the conditional probability that participant $i$ is assigned treatment $A$ versus treatment $C$ given measured covariates $\bm{x}_i$ \cite{rosenbaum1983central}. 

Having fitted the model in Equation \ref{equation7}, e.g.~using maximum-likelihood estimation, propensity scores for the subjects in the index trial are predicted using:
\begin{equation}
\hat{e}_i = \expit[\hat{\beta}_0 + \bm{x}_i \hat{\bm{\beta}}_{\bm{1}}],   
\label{equation8}
\end{equation}
where $\expit(\cdot)=\exp(\cdot)/[1+\exp(\cdot)]$, $\hat{\beta}_0$ and $\hat{\bm{\beta}}_{\bm{1}}$ are point estimates of the logistic regression parameters, and $\hat{e}_i$ is an estimate of $e_i$. Inverse probability of treatment weights are constructed by taking the reciprocal of the estimated conditional probability of the treatment assigned in the index study \cite{lunceford2004stratification}. That would be $1/\hat{e}_i$ for units under treatment $A$ and $1/(1-\hat{e}_i)$ for units under treatment $C$. 

Consequently, the weights produced by the standard MAIC (Equation \ref{equation4}) are rescaled by the estimated inverse probability of treatment weights. The contribution of each subject $i$ in the IPD is weighted by:

\begin{equation}
\hat{\omega}_i = \frac{t_i \hat{w}_i}{\hat{e}_i} + \frac{(1-t_i)  \hat{w}_i}{(1-\hat{e}_i)}.
\label{equation9}
\end{equation}
The weights $\{ \hat{w}_i, \hspace{0.1cm} i=1,\dots,n \}$ estimated by the standard MAIC are odds, constrained to be positive. These balance the index and competitor study studies in terms of the selected effect modifier moments. The estimated propensity scores $\{ \hat{e}_i, \hspace{0.1cm} i=1,\dots,n \}$ are probabilities bounded away from zero and one. Therefore, the weights $\{ \hat{\omega}_i, \hspace{0.1cm} i=1,\dots,n \}$ produced by 2SMAIC in Equation \ref{equation9} are constrained to be positive. These weights achieve balance in effect modifier moments across studies, but also seek to balance covariate moments between the index trial's treatment groups. 

Marginal mean outcomes for treatments $A$ and $C$ in the competitor study sample are estimated as the weighted average of observed outcomes:
\begin{equation}
\hat{\mu}^{(2)}_t = \frac{\sum_{i=1}^{n_t} y_{i,t} \hat{\omega}_{i,t}}{\sum_{i=1}^{n_t} \hat{\omega}_{i,t}},
\label{equation10}
\end{equation}
where $\hat{\omega}_{i,t}$ is the weight assigned to patient $i$ under treatment $t$. One can convert the mean absolute outcome predictions generated by Equation \ref{equation10} to the linear predictor scale, and compute the marginal treatment effect for $A$ vs.~$C$ in $S=2$ as the difference between the average linear predictions, as per Equation \ref{equation6}. Alternatively, a weighted regression of outcome on treatment alone can be fitted to the IPD, in which case the treatment coefficient of the fitted model represents the estimated marginal treatment effect $\hat{\Delta}_{10}^{(2)}$ for $A$ vs.~$C$ in $S=2$. 

Inference can be based on a robust sandwich-type variance estimator or on resampling approaches such as the non-parametric bootstrap. As noted previously, the sandwich variance estimator is biased downwards when the ESS after weighting is small, leading to overprecision. In practice, the non-parametric bootstrap is a preferred option, re-estimating both the trial assignment model and the treatment assignment model in each iteration. This approach explicitly accounts for the estimation of the weights and is expected to perform better where the ESS is small. 

It may seem counter-intuitive to estimate the treatment assignment mechanism when the index trial is an RCT. The randomized design implies that the true propensity scores $\{ e_i, \hspace{0.1cm} i=1,\dots, n \}$ are fixed and known. For instance, consider a marginally randomized two-arm trial with a 1:1 treatment allocation ratio. The trial investigators have determined in advance that the probability of being assigned to active treatment versus control is $e_i = 0.5$ for all $i$. 

The rationale for estimating the propensity scores is the following. Randomization guarantees that there is no confounding on expectation \cite{senn1994testing}. Nevertheless, covariate balance is a large-sample property, and one may still observe residual covariate imbalances between treatment groups due to chance, especially when the trial sample size is small \cite{li2020rerandomization}. As formulated by Senn \cite{senn1994testing}, ``over all randomizations the groups are balanced; for a particular randomization they are unbalanced.'' The use of estimated propensity scores allows to correct for random finite-sample imbalances in prognostic baseline covariates. In the RCT literature, inverse probability of treatment weighting is an established approach for covariate adjustment \cite{morris2022planning}, and has increased precision, efficiency and power with respect to unadjusted analyses in the estimation of marginal treatment effects \cite{raad2020evaluation, williamson2014variance}. 

Insofar, the use of anchored MAIC has been limited to situations where the index trial is an RCT. 2SMAIC can be used when the index study is observational, provided that the baseline covariates in $\bm{x}$ offer sufficient control for confounding. In non-randomized studies, the true propensity score for each participant in the index study is unknown, and additional conditions are needed to produce internally valid estimates of the marginal treatment effect for $A$ vs.~$C$. These are: (1) conditional exchangeability over treatment assignment \cite{hernan2006estimating}; and (2) positivity of treatment assignment \cite{westreich2010invited}. Randomized trials tend to meet these assumptions by design. The assumptions have conceptual parallels with the conditional transportability and overlap conditions previously described for MAIC. 

The first assumption indicates that the potential outcomes of subjects in each treatment group are independent of the treatment assigned after conditioning on the selected covariates. It relies on all confounders of the effect of treatment on outcome being measured and accounted for \cite{holland1986statistics}. The second assumption indicates that, for every participant in the index study, the probability of being assigned to either treatment is positive, conditional on the covariates selected to ensure exchangeability \cite{westreich2010invited}. This requires overlap between the joint covariate distributions of the subjects under treatment $A$ and under treatment $C$. This assumption is threatened if there are few or no individuals from either treatment group in certain covariate subgroups/strata.

\section*{Simulation study}

\subsection*{Aims}

The objectives of the simulation study are to provide proof-of-principle for 2SMAIC and to benchmark its statistical performance against that of MAIC in an anchored setting where the index study is an RCT. We also investigate whether weight truncation can improve the performance of MAIC and 2SMAIC by reducing the variance caused by extreme weights. 

Each method is assessed using the following frequentist characteristics \cite{morris2019using}: (1) unbiasedness; (2) precision; (3) efficiency (accuracy); and (4) randomization validity (valid confidence interval estimates). The selected performance metrics specifically evaluate these criteria. The ADEMP (Aims, Data-generating mechanisms, Estimands, Methods, Performance measures) framework \cite{morris2019using} is used to describe the simulation study design. Example \texttt{R} code implementing the methodologies is provided in Additional file 1. All simulations and analyses have been conducted in \texttt{R} software version 4.1.1 \cite{team2013r}.\footnote{The files required to run the simulations are available at \url{http://github.com/remiroazocar/Maic2stage}.}

\subsection*{Data-generating mechanisms}

We consider continuous outcomes using the mean difference as the measure of effect. For the index and competitor studies, outcome $y_i$ for participant $i$ is generated as:
\begin{equation*}
y_i = \beta_0 + \bm{x}_i\bm{\beta_1} +  \left (\beta_t + \bm{x}_i\bm{\beta_2} \right)\mathds{1}(t_i=1) + \epsilon_i,
\end{equation*}
using the notation of the index study data. Each $\bm{x}_i$ contains the values of three correlated continuous covariates, which have been simulated from a multivariate normal distribution with pre-specified means and covariance matrix. There is some positive correlation between the three covariates, with pairwise Pearson correlation levels set to 0.2. The covariates have main effects and are prognostic of individual-level outcomes independently of treatment. They also have first-order covariate-treatment product terms, thereby modifying the conditional (and marginal) effects of both $A$ and $B$ versus $C$ on the mean difference scale, i.e., $\bm{z}$ is equivalent to $\bm{x}$. The term $\epsilon_i$ is an error term for subject $i$ generated from a standard (zero-mean, unit-variance) normal distribution. 

The main ``prognostic'' coefficient $\beta_{1,k}=2$ for each covariate $k$. This is considered a strong covariate-outcome association. The interaction coefficient $\beta_{2,k}=1$ for each covariate $k$, indicating notable effect modification. We set the intercept $\beta_0=5$. Active treatments $A$ and $B$ are assumed to have the same set of effect modifiers with respect to the common comparator, and identical interaction coefficients for each effect modifier. Consequently, the shared (conditional) effect modifier assumption holds \cite{phillippo2016nice}. The main treatment coefficient $\beta_t=-2$ is considered a strong conditional treatment effect versus the control at baseline (when the covariate values are zero).  

The continuous outcome may represent a biomarker indicating disease severity. The covariates are comorbidities associated with higher values of the biomarker and which interact with the active treatments to hinder their effect versus the control. 

It is assumed that the index and competitor studies are simple, marginally randomized, RCTs. The number of participants in the competitor RCT is 300, with a 1:1 allocation ratio for active treatment vs.~control. For this study, individual-level covariates are summarized as means. These would be available to the analyst in a table of baseline characteristics in the study publication. Individual-level outcomes are aggregated by fitting a simple linear regression of outcome on treatment to produce an unadjusted estimate of the marginal mean difference for $B$ vs.~$C$, with its corresponding nominal standard error. This information would also be available in the published study. 

We adopt a factorial arrangement using two index trial sample sizes times three overlap settings. This results in a total of six simulation scenarios. The following parameter values are varied:

\begin{itemize}
    \item Sample sizes of $n \in \{ 140, 200\}$ are considered for the index trial, with an allocation ratio of 1:1 for intervention $A$ vs.~$C$. The sample sizes are small but not unusual in applications of MAIC in HTA submissions \cite{phillippo2019population}. It is anticipated that smaller trials are subject to a greater chance of covariate imbalance than larger trials \cite{thompson2015covariate}.   
    \item The level of (deterministic) covariate overlap. Covariates follow normal marginal distributions in both studies. For the competitor trial, the marginal distribution means are fixed at 0.6. For the index trial, the mean $\mu_k \in \{0.5, 0.4, 0.3\}$ for each covariate $k$. These settings yield strong, moderate and poor overlap, respectively. The standard deviations in both studies are fixed at 0.4, i.e., a one standard deviation increase in each covariate is associated with a 0.8 unit increase in the outcome. Greater covariate imbalances across studies lead to poorer overlap between the trials' target populations, which translates into more variable weights and a lower ESS. Unless otherwise stated, when describing the results of the simulation study, ``covariate overlap'' relates to deterministic overlap between the trials' target populations and not to random violations arising due to small sample sizes. 
\end{itemize}

\subsection*{Estimands}

The target estimand is the marginal mean difference for $A$ vs.~$B$ in $S=2$. The treatment coefficient $\beta_t=-2$ is the same for both $A$ vs.~$C$ and $B$ vs.~$C$, and the shared (conditional) effect modifier assumption holds. Therefore, the true conditional treatment effects for $A$ vs.~$C$ and $B$ vs.~$C$ in $S=2$ are the same ($-2+3\times(0.6\times1)=-0.2$). Because mean differences are collapsible, the true marginal treatment effects for $A$ vs.~$C$ and $B$ vs.~$C$ coincide with the corresponding conditional estimands. The true marginal effect for $A$ vs.~$B$ in $S=2$ is a composite of that for $A$ vs.~$C$ and $B$ vs.~$C$, which cancel out. Consequently, the true marginal mean difference for $A$ vs.~$B$ in $S=2$ is zero. 

Note that all the methods being compared conduct the same unadjusted analysis to estimate the marginal treatment effect of $B$ vs.~$C$. Because the competitor study is a randomized trial, this estimate should be unbiased with respect to the corresponding marginal estimand in $S=2$. Therefore, differences in performance between the methods will arise from the comparison between $A$ and $C$, for which marginal and conditional estimands are non-null. 

\subsection*{Methods}

Each simulated dataset is analyzed using the following methods:

\begin{itemize}
\item Matching-adjusted indirect comparison (MAIC). The trial assignment model in Equation \ref{equation2} contains main effect terms for all three effect modifiers --- only covariate means are balanced. The objective function in Equation \ref{equation3} is minimized using BFGS \cite{nocedal2006numerical}. The weights estimated by Equation \ref{equation4} are used to fit a weighted simple linear regression of outcome on treatment to the index trial IPD. 

\item Two-stage matching-adjusted indirect comparison (2SMAIC). We follow the same steps as for the standard MAIC. In addition, the treatment assignment model in Equation \ref{equation7} is fitted to the index study IPD, including main effect terms for all three baseline covariates. Propensity score estimates are generated by Equation \ref{equation8} and combined with the weights generated by Equation \ref{equation4} as per Equation \ref{equation9}. The resulting weights are used to fit a weighted simple linear regression of outcome on treatment to the index trial IPD. 

\item Truncated matching-adjusted indirect comparison (T-MAIC). This approach is identical to MAIC but the highest estimated weights (Equation \ref{equation4}) are truncated using a 95th percentile cutpoint, following Susukida et al.~\cite{susukida2018comparing, susukida2018generalizability}, Webster-Clark et al.~\cite{webster2019diagnostic}, and Lee et al.~\cite{lee2011weight}. Specifically, all weights above the 95th percentile are replaced by the value of the 95th percentile. 

\item Truncated two-stage matching-adjusted indirect comparison (T-2SMAIC). This approach is identical to 2SMAIC but all the estimated weights (Equation \ref{equation9}) larger than the 95th percentile are set equal to the 95th percentile. 
\end{itemize}

All approaches use the ordinary non-parametric bootstrap to estimate the variance of the $A$ vs.~$C$ marginal treatment effect. 2,000 resamples of each simulated dataset are drawn with replacement \cite{efron1994introduction, carpenter2000bootstrap}. Due to patient-level data limitations for the competitor study, only the IPD of the index trial are resampled in the implementation of the bootstrap. The average marginal mean difference for $A$ vs.~$C$ in $S=2$ is computed as the average across the bootstrap resamples. Its standard error is the standard deviation across these resamples. For the ``one-stage'' MAIC approaches, each bootstrap iteration re-estimates the trial assignment model. For the ``two-stage'' MAIC approaches, both the trial assignment and the treatment assignment model are re-estimated in each iteration. 

All methods perform the indirect treatment comparison in a final stage, where the results of the study-specific analyses are combined. The marginal mean difference for $A$ vs.~$B$ is obtained by directly substituting the point estimates $\hat{\Delta}_{10}^{(2)}$ and $\hat{\Delta}_{20}^{(2)}$ in Equation \ref{equation1}. Its variance is estimated by adding the point estimates of the variance for the within-study treatment effect estimates. Wald-type 95\% confidence interval estimates are constructed using normal distributions.

\subsection*{Performance measures}

We generate 5,000 simulated datasets per simulation scenario. For each scenario and analysis method, the following performance metrics are computed over the 5,000 replicates: (1) bias in the estimated treatment effect; (2) empirical standard error (ESE); (3) mean square error (MSE); and (4) empirical coverage rate of the 95\% confidence interval estimates. These metrics are defined explicitly in prior work \cite{remiro2021methods, morris2019using}. 

The bias evaluates aim 1 of the simulation study. It is equal to the average treatment effect estimate across the simulations because the true estimand is zero ($\Delta_{12}^{(2)}=0$). The ESE targets aim 2 and is the standard deviation of the treatment effect estimates over the 5,000 runs. The MSE represents the average squared bias plus the variance across the simulated replicates. It measures overall efficiency (aim 3), accounting for both bias (aim 1) and precision (aim 2). Coverage assesses aim 4, and is computed as the percentage of estimated 95\% confidence intervals that contain the true value of the estimand. 

We have used 5,000 replicates per scenario based on the analysis method and scenario with the largest long-run variability (standard MAIC with $n=140$ and poor overlap). Assuming $\textnormal{SD}(\hat{\Delta}_{12}^{(2)}) \leq 0.53$, the Monte Carlo standard error (MCSE) of the bias is at most $\sqrt{\textnormal{Var}(\hat{\Delta}_{12}^{(2)})/N_{sim}}=\sqrt{0.28/5000}=0.007$ under 5,000 simulations per scenario, and the MCSE of the coverage, based on an empirical coverage rate of $95\%$ is $\left(\sqrt{(95 \times 5)/5000}\right)\%=0.31\%$, with the worst-case being $0.71\%$ under $50\%$ coverage. These are considered adequate levels of simulation uncertainty.  

\section*{Results}

Performance measures for all methods and simulation scenarios are reported in Figure \ref{fig1}. The strong overlap settings are at the top (in ascending order of index trial sample size), followed by the moderate overlap settings and the poor overlap settings at the bottom. For each data-generating mechanism, there is a ridgeline plot visualizing the spread of point estimates for the marginal $A$ vs.~$B$ treatment effect over the 5,000 simulation replicates. Below each plot, a table summarizing the performance metrics of each method is displayed. MCSEs for each metric, used to quantify the simulation uncertainty, have been computed and are presented in parentheses alongside the average of each performance measure. These are considered negligible due to the large number of simulated datasets per scenario. In Figure \ref{fig1}, Cov denotes the empirical coverage rate of the 95\% confidence interval estimates. 

In the most extreme scenario ($n=140$ and poor covariate overlap), weights could not be estimated for 1 of the 5,000 simulated datasets. This was due to total separation: empirically, all the values observed in the index trial for one of the baseline covariates were below the competitor study mean. Therefore, there were no feasible solutions minimizing the objective function in Equation \ref{equation3}. The affected replicate was discarded, and 4,999 simulated datasets were analyzed in the corresponding scenario. With respect to the treatment assignment model, empirical overlap between treatment arms was always excellent due to randomization in the index trial. 

\subsection*{Bias}

Even with the small index trial sample sizes, bias is similarly low for MAIC and 2SMAIC without truncation in all simulation scenarios. There is a slight increase in bias as the ESS after weighting decreases, with the bias of highest magnitude occurring with $n=140$ and poor covariate overlap (the scenario with the lowest ESS after weighting) for MAIC (-0.041) and 2SMAIC (-0.031). In absolute terms, the bias of 2SMAIC is smaller than that of MAIC in all simulation scenarios. For 2SMAIC, it is within Monte Carlo error of zero in all scenarios except in the most extreme setting, mentioned earlier, and in the setting with $n=200$ and moderate overlap (-0.008). Of all methods, 2SMAIC produces the lowest bias in every simulation scenario. 

Weight truncation increases absolute bias in all scenarios. T-MAIC and T-2SMAIC consistently exhibit greater bias than MAIC and 2SMAIC. When overlap is strong, truncation only induces bias very slightly. As overlap is reduced, the bias induced by truncation is more noticeable, particularly in the $n=140$ settings. For instance, the bias for T-MAIC and T-2SMAIC in the scenarios with poor overlap is substantial (for $n=140$: 0.157 and 0.160, respectively; for $n=200$, 0.149 and 0.153). For the truncated methods, the magnitude of the bias also appears to increase as the ESS after weighting decreases. 

\subsection*{Precision}

As expected, all methods incur precision losses as the number of subjects in the index trial and covariate overlap decrease. Despite enforcing randomization in the index trial, 2SMAIC increases precision, as measured by the ESE, with respect to MAIC in every simulation scenario. Reductions in ESE are more dramatic in the $n=140$ settings than in the $n=200$ settings. This is attributed to a greater chance of empirical covariate imbalances with smaller sample sizes. Interestingly, reduced covariate overlap seems to minimize the effect of incorporating the second (treatment assignment) stage. This is likely due to precision gains being offset by the presence of extreme weights, which lead to large reductions in ESS and inflate the ESE. The same trends are revealed for T-2SMAIC with respect to T-MAIC across the simulation scenarios. Both ``two-stage'' versions have reduced ESEs compared to their ``one-stage'' counterparts in all scenarios.  

Weight truncation decreases the ESE across all simulation scenarios for one-stage and two-stage MAIC. This is to be expected as the influence of outlying weights is reduced. When overlap is strong, truncation offers only a small improvement in precision. This has little impact in comparison to the inclusion of a second stage in MAIC. For instance, under strong overlap and $n=140$, the ESE for MAIC and 2SMAIC is 0.516 and 0.386, respectively; compared to ESEs of 0.489 and 0.371 for the corresponding truncated versions. 

The precision gains of weight truncation become more considerable as overlap weakens and the extremity of the weights increases. When overlap is poor, truncation reduces the ESE more sharply than the incorporation of a second stage in MAIC. For example, under poor overlap and $n=140$, the ESE of MAIC and 2SMAIC is 0.767 and 0.703, respectively, and that of the truncated versions is 0.563 and 0.519. Unsurprisingly, the combination of incorporating the second stage and truncating the weights is most effective at variance reduction. As $n$ decreases, precision seems to be more markedly reduced for the one-stage approaches than for the two-stage approaches, and for the untruncated approaches than for the truncated ones.  

Where covariate overlap is strong, T-2SMAIC has the highest precision, followed by 2SMAIC, T-MAIC and MAIC. Where covariate overlap is moderate or poor, T-2SMAIC has the highest precision, followed by T-MAIC, 2SMAIC and MAIC. 

\subsection*{Efficiency}

As per the ESE, MSE values decrease for all methods as the index trial sample size and covariate overlap increase. In agreement with the trends for precision, the two-stage versions of MAIC increase efficiency with respect to the corresponding one-stage methods in all scenarios, particularly in the $n=140$ settings. Efficiency gains for the two-stage approaches are stronger where covariate overlap is strong and become less noticeable as covariate overlap weakens, due to extreme weights. For instance, with strong overlap and $n=200$, MSEs for MAIC and 2SMAIC are 0.205 and 0.127, respectively. With poor overlap and $n=200$, these are 0.459 and 0.393, respectively.  

Differences in MSE between methods are driven more by comparative precision than bias. This is expected in the strong overlap scenarios, where the bias for all methods is negligible, but also occurs in the poor overlap scenarios. The precision gains of truncation more than counterbalance the increase in bias when the variability of the weights is high. As overlap decreases, the relative efficiency of the truncated versus the untruncated approaches is markedly improved. For example, with poor overlap and $n=200$, the MSE of T-MAIC and T-2SMAIC is 0.263 and 0.233, respectively (compared to MSEs of 0.459 and 0.393 for MAIC and 2SMAIC). 

T-2SMAIC is the most efficient method and MAIC is the least efficient method across all simulation scenarios in terms of MSE. Where covariate overlap is strong, T-2SMAIC yields the highest efficiency, followed by 2SMAIC, T-MAIC and MAIC. Where overlap is poor, T-2SMAIC has the highest efficiency, followed by T-MAIC, 2SMAIC and MAIC. Where overlap is moderate, 2SMAIC and T-MAIC have comparable efficiency. 

\subsection*{Coverage}

From a frequentist perspective, 95\% confidence interval estimates should include the true estimand 95\% of the time. Namely, empirical coverage rates should equal the nominal coverage rates to ensure appropriate type I error rates for testing a ``no effect'' null hypothesis. Theoretically, due to our use of 5,000 Monte Carlo simulations per scenario, empirical coverage rates are statistically significantly different to the desired 0.95 if they are under 0.944 or over 0.956. 

Empirical coverage rates for MAIC are statistically significantly different to the nominal coverage rate in all but one scenario: that with strong overlap and $n=200$. Where covariate overlap is strong or moderate, all other methods exhibit empirical coverage rates that are very close to the advertised nominal values (all differences are not significantly different, except for T-MAIC in the scenario with strong overlap and $n=140$). 

There is discernible undercoverage for all methods when overlap is poor. This is particularly the case for the approaches without truncation. For instance, for the smallest sample size ($n=140$) with poor overlap, the empirical coverage rate is 0.900 for MAIC and 0.917 for 2SMAIC. These anti-conservative inferences could arise from the use of normal distribution-based confidence intervals when the ESS after weighting is small. While the large-sample normal approximation produces asymptotically valid inferences, a reasonable alternative in small ESS scenarios could be the use of a t-distribution. An open question is how to choose the degrees of freedom of the t-distribution.  

Interestingly, coverage drops are larger for the untruncated approaches than for the truncated approaches as overlap weakens. This is surprising because the truncated methods induce sizeable bias in the poor overlap settings, and one would have expected coverage rates to be degraded further by this bias. Weight truncation has improved coverage rates in another simulation study in a different context \cite{lee2011weight}. This warrants further investigation. Overcoverage is not a problem for any of the methods as the empirical coverage rates never rise above 0.956.    

\section*{Discussion}

\subsection*{Limitations of simulation study}

In all simulation scenarios, two-stage methods offer enhanced precision and efficiency with respect to one-stage methods. These gains are likely linked to the prognostic strength of the baseline covariates included in the treatment assignment model. We have assumed, as is typically the case in practice, that the baseline covariates are prognostic of outcome. Less notable increases in precision and efficiency are expected when covariate-outcome associations are lower.  

All approaches depend on the critical assumption of conditional transportability over trials. Given the somewhat arbitrary and unclear process driving selection into different studies in our context (in reality, there is not a formal assignment process determining whether subjects are in study sample $S=1$ or $S=2$), I have not specified a true trial assignment mechanism in the simulation study. Nevertheless, the true outcome-generating mechanism imposes linearity and additivity assumptions in the covariate-outcome associations and the treatment-by-covariate interactions. Conditional transportability holds because the trial assignment model balances means for all the covariates that modify the marginal treatment effect of $A$ vs.~$C$. 

In real-life scenarios, it is entirely possible that more complex relationships underlie the outcome-generating process. These would potentially require balancing higher-order moments, covariate-by-covariate interactions and non-linear transformations of the covariates. In practice, sensitivity analyses will be required to explore whether there are discrepancies in the results produced by different model specifications.

The methods evaluated in this article focus on correcting for imbalances in baseline covariates, i.e., the `P' in the PICO (Population, Intervention, Comparator, Outcome) framework \cite{richardson1995well}. Nevertheless, there are other kinds of differences which may bias indirect treatment comparisons, e.g.~in comparator or endpoint definitions. The methodologies that have been evaluated in this article cannot adjust for these types of differences. 

\subsection*{Contributions in light of recent simulation studies}

Prior simulation studies in the context of anchored indirect treatment comparisons have concluded that outcome regression is more precise and efficient than weighting when the conditional outcome-generating mechanism is known \cite{remiro2021parametric, remiro2020marginalization}. This is likely to remain the case despite the performance gains of 2SMAIC and the truncated approaches with respect to MAIC. 

Nevertheless, there is one caveat. In these studies, the (one-stage) MAIC trial assignment model only accounts for covariates that are marginal effect modifiers. The reason for this is that including prognostic covariates that are not effect modifiers deteriorates precision without improving the potential for bias reduction. Conversely, the outcome regression approaches have included all prognostic covariates in the outcome model, making use of this prognostic information to increase precision and efficiency. Therefore, the equipoise or fairness in previous comparisons between weighting and outcome regression is debatable.

With 2SMAIC, weighting approaches can now make use of this prognostic information by including the relevant covariates in the treatment assignment model. Future simulation studies comparing weighting and outcome regression should involve 2SMAIC as opposed to its one-stage counterpart, particularly in these ``perfect information'' scenarios.  

\subsection*{Extension to observational studies}

Almost invariably, anchored MAIC has been applied in a setting where the index trial is randomized. In this setting, the inclusion of the treatment assignment model leads to efficiency gains by increasing precision. Any reduction in bias will be, at most, modest due to the internal validity of the index trial. Nevertheless, in situations where the index study is observational, the treatment assignment model can be useful to reduce internal validity bias due to confounding.

Transporting the results of a non-randomized study from $S=1$ to $S=2$ requires further untestable assumptions. Additional barriers are: (1) susceptibility to unmeasured confounding; and (2) positivity issues. Due to randomization, there is typically excellent overlap between treatment arms in RCTs. However, theoretical (deterministic) violations of positivity may occur in observational study designs \cite{leger2022causal, westreich2010invited, petersen2012diagnosing}, e.g.~subjects with certain covariate values may have a contraindication for receiving one of the treatments, resulting in a null probability of treatment assignment. 

In addition to these conceptual problems, ``chance'' violations of positivity may occur with small sample sizes or high-dimensional data due to sampling variability, in both randomized and non-randomized studies. These have not been observed in this simulation study. Near-violations of positivity between treatment arms may lead to extreme inverse probability of treatment weights \cite{li2019addressing}, further inflating variance in 2SMAIC.

Finally, it is worth noting that observational study designs have traditionally been more prone than RCTs to additional causes of internal validity bias, e.g.~missing outcome data, measurement error or protocol deviations \cite{deeks2003evaluating}.

\subsection*{Approaches for variance reduction}

Weight truncation is a relatively informal but easily implemented method to improve precision by restricting the contribution of extreme weights. The choice of a 95th percentile cutoff is based on prior literature and is somewhat arbitrary, but worked well in this simulation study. Alternative threshold values could be considered. 

Lower thresholds will further reduce variance at the cost of introducing more bias and shifting the target population or estimand definition further \cite{cole2008constructing, xiao2013comparison}. The ideal truncation level will vary on a case-by-case basis and can be set empirically, e.g. by progressively truncating the weights \cite{cole2008constructing, kish1992weighting}. Density plots are likely helpful to assess the dispersion of the weights and identify an optimal cutoff point. Weight truncation is likely of little utility where there is sufficient overlap and the weights are well-behaved. Efficiency gains are expected to decrease with larger sample sizes, as the induced bias could potentially offset the reduction of variance. 

We have only explored two strategies to improve efficiency: (1) modeling the trial assignment mechanism; and (2) truncating the weights that are above a certain level. Nevertheless, there are other approaches that could be used in practical applications, either on their own or combined with the procedures explored in this article. Weight trimming \cite{crump2009dealing} is closely related to weight truncation. It involves excluding the subjects with outlying weights, thereby sharing many of the limitations of truncation: setting arbitrary cutoff points, and changing the target population even further. Trimming is unappealing because it directly throws away information, discarding data from some individuals, and likely losing precision with respect to truncation. 

The use of stabilized weights is often recommended to gain precision and efficiency \cite{cole2008constructing, austin2015moving}, particularly when the weights are highly variable. In the implementations of MAIC in this article, the fitted weighted outcome model is considered to be ``saturated'' (i.e., cannot be misspecified) because it is a marginal model of outcome on a time-fixed binary treatment \cite{shiba2021using}. For saturated models, stabilized and unstabilized weights give identical results \cite{shiba2021using}. Nevertheless, weight stabilization is encouraged when the weighted outcome model is unsaturated, e.g.~with dynamic (time-varying) or continuous-valued treatment regimens \cite{robins2000marginal, robins2009estimation}. 

Another approach that has been used to gain efficiency is overlap weighting \cite{zeng2021propensity, desai2019alternative}. It also changes the target estimand, estimating treatment effects in a subsample with good overlap. While the approach is worth consideration, it is challenging to implement in our context because IPD are unavailable for the competitor study. 

In the Background section, I referred to the weight estimation procedure by Jackson et al.~\cite{jackson2020alternative}, which satisfies the method of moments while maximizing the ESS, thereby reducing the dispersion of the weights. 2SMAIC is a modular framework and this approach could be used instead of the standard method of moments to estimate the trial assignment odds weights. Different weighting modules could be incorporated to account for missing outcomes \cite{seaman2013review}, treatment switching \cite{robins2000correcting, latimer2017adjusting} and other forms of non-adherence to the protocol \cite{cain2009inverse} in the index trial.

\section*{Conclusions}

I have introduced 2SMAIC, an extension of MAIC that combines a model for the treatment assignment mechanism in the index trial with a model for the trial assignment mechanism. The first model accounts for covariate differences between treatment arms, producing inverse probability weights that can balance the treatment groups of the index study. The second model accounts for effect modifier differences between studies, generating odds weights that achieve balance across trials and allow us to transport the marginal effect for $A$ vs.~$C$ from $S=1$ to $S=2$. In 2SMAIC, both weights are combined to attain balance between the treatment arms of the index trial and across the studies.

The statistical performance of 2SMAIC has been investigated in scenarios where the index study is an RCT. We find that the addition of a second (treatment assignment) stage increases precision and efficiency with respect to the standard one-stage MAIC. It does so without inducing bias and being less prone to undercoverage. Efficiency and precision gains are prominent when the index trial has a small sample size, in which case it is subject to empirical imbalances in prognostic baseline covariates. Two-stage MAIC accounts for these chance imbalances through the treatment assignment model, mitigating the precision loss coming with decreasing sample sizes. Precision and efficiency gains are attenuated when there is poor overlap between the target populations of the studies, due to the high extremity of the estimated weights. 

The inclusion of weight truncation approaches has been evaluated for the first time in the context of MAIC. The one-stage and two-stage approaches produced very little bias before truncation was applied. Where covariate overlap was strong and the variability of the weights tolerable, truncation only improved precision and efficiency slightly, while inducing bias. The benefits of truncation become more apparent in situations with weakening overlap, where it diminishes the influence of extreme weights, substantially improving precision and even coverage with respect to the untruncated approaches. 

Due to bias-variance trade-offs, precision improvements always come at the cost of bias. In this simulation study, the trade-off favors variance reduction over the induced bias, with truncation improving efficiency in all scenarios. Nevertheless, truncation is likely unnecessary when the weights are well-behaved and the ESS after weighting is sizeable. The combination of a second stage and weight truncation is most effective in improving precision and efficiency in all simulation scenarios. 

When covariate overlap is poor, undercoverage is an issue for all methods, particularly for the untruncated approaches. Novel outcome regression-based techniques \cite{phillippo2020assessing, remiro2021parametric, remiro2020marginalization, phillippo2021target, phillippo2020multilevel} may be preferable in these situations. The development of doubly robust approaches that combine outcome modeling with a model for the trial assignment weights is also attractive, as these would give researchers two chances for correct model specification. 

In the absence of a common comparator group, unanchored comparisons contrast the outcomes of single treatment arms between studies. Because one of the stages relies on estimating the treatment assignment mechanism in the index study, the two-stage approaches are not applicable in the unanchored case. This is a limitation, as many applications of covariate-adjusted indirect comparisons are in this setting \cite{phillippo2019population}, both in published studies and in health technology appraisals.

Finally, I address a misconception that has arisen recently in the literature \cite{phillippo2021target, remiro2022target}. It is believed that MAIC replicates the unadjusted analysis that would be performed in a hypothetical ``ideal RCT'' because it targets a marginal estimand, and that MAIC cannot make use of information on prognostic covariates. While all approaches to MAIC target marginal estimands, these produce covariate-adjusted estimates of the marginal effect. The standard one-stage approach to MAIC accounts for covariate differences across studies. The two-stage approaches introduced in this article generate covariate-adjusted estimates that also account for imbalances between treatment arms in the index trial, as is the case in covariate-adjusted analyses of RCTs.   


\begin{backmatter}

\section*{Declarations}

\vspace{0.3cm}

\section*{Acknowledgments}

Not applicable.

\section*{Funding}

No financial support was provided for this research.

\section*{Author affiliations}

Medical Affairs Statistics, Bayer plc, Reading, United Kingdom. 

Department of Statistical Science, University College London, London, United Kingdom. 

\section*{Author contributions}

ARA conceived the research idea, developed the methodology, performed the analyses, prepared the figures, and wrote and reviewed the manuscript.

\section*{Abbreviations}

2SMAIC - Two-stage matching-adjusted indirect comparison \\
ALD - Aggregate-level data \\
ESE - Empirical standard error \\
ESS - Effective sample size \\
IPD - Individual patient data \\
HTA - Health technology assessment \\
MAIC - Matching-adjusted indirect comparison \\
MCSE - Monte Carlo standard error \\
MSE - Mean square error \\
PICO - Population, Intervention, Comparator, Outcome \\
RCT - Randomized controlled trial \\
T-MAIC - Truncated matching-adjusted indirect comparison \\
T-2SMAIC - Truncated two-stage matching-adjusted indirect comparison \\

\section*{Availability of data and materials}

The files required to generate the data, run the simulations, and reproduce the results are available at \url{http://github.com/remiroazocar/Maic2stage}.

\section*{Ethics approval and consent to participate}

Not applicable.

\section*{Competing interests}
  
ARA is employed by Bayer plc. The author declares no conflicts of interest as this research is purely methodological.

\section*{Consent for publication}

Not applicable.


\bibliographystyle{bmc-mathphys} 
\bibliography{bmc_article}      




\clearpage

\section*{Figures}
\begin{figure}[!htb]
  \includegraphics[width=\textwidth]{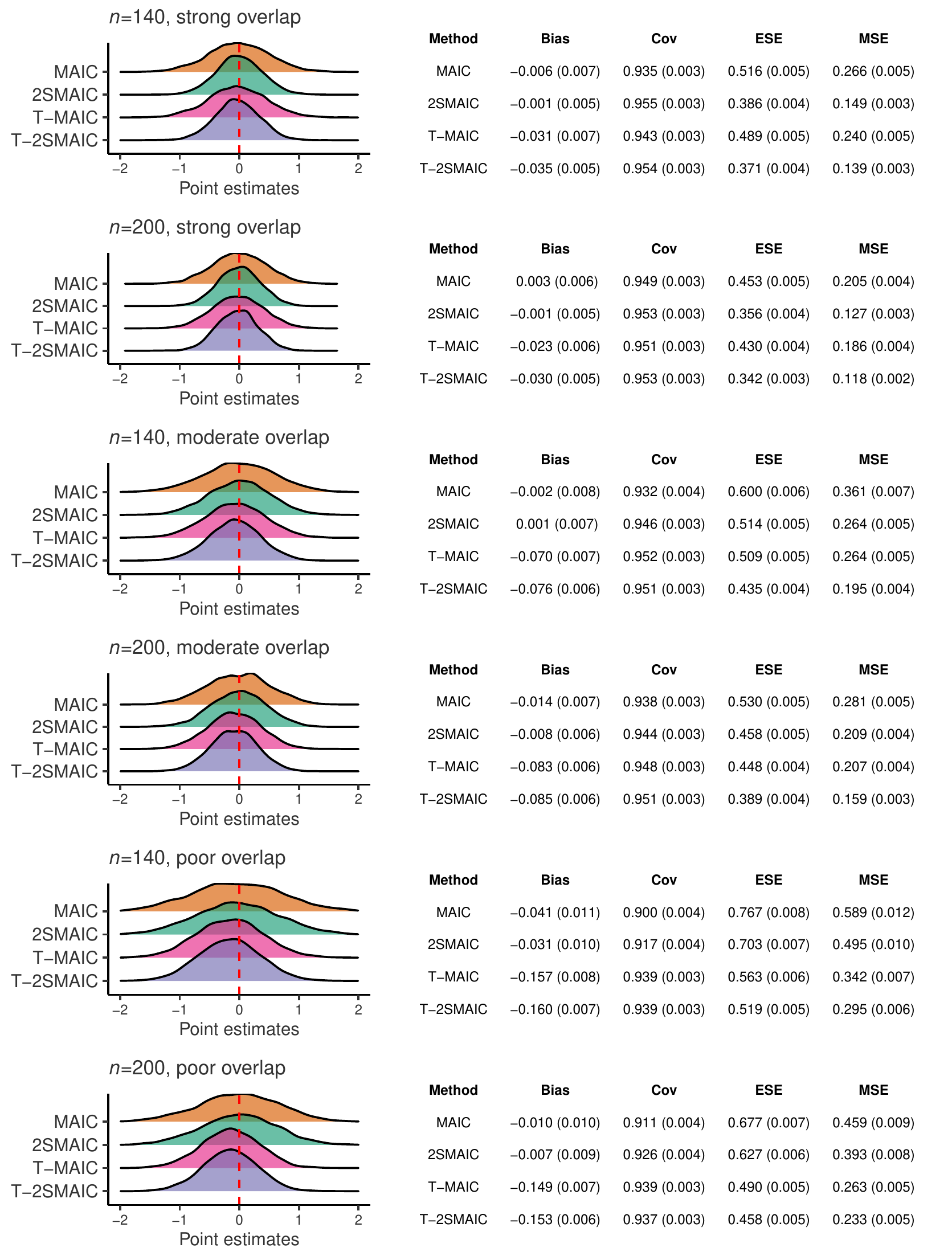}
  \caption{\csentence{Simulation study results.}
      Point estimates of the treatment effect and performance metrics for all methods and simulation scenarios.}
     \label{fig1}
\end{figure}




\section*{Additional Files}
  \subsection*{Additional file 1 --- Supplementary Material}
    \texttt{AdditionalFile1.pdf} provides example \texttt{R} code implementing the methodologies evaluated in the simulation study.
\end{backmatter}
\end{document}